\documentclass{elsart}

\pdfoutput=1

\usepackage{amssymb,graphicx,url, amsmath, rotating}
\usepackage{wasysym}
\usepackage[scriptsize]{subfigure}

\begin{document}

\begin{frontmatter}



\title{On localizations in minimal cellular automata model of two-species mutualism}



\author{Andrew Adamatzky$^1$ and Martin Grube$^2$}

\address{$^1$ University of the West of England Bristol BS16 1QY United Kingdom\\ 
andrew.adamatzky@uwe.ac.uk\\
         $^2$ Institut f\"{u}r Pflanzenwissenschaften Karl-Franzens-Universit\"{a}t Graz, Austria\\ 
martin.grube@kfunigraz.ac.at
}

\begin{abstract}
A mutualism is an interaction where the involved species benefit from each other. We study a two-dimensional hexagonal 
three-state cellular automaton model of a two-species mutualistic system. The simple model is characterized by four 
parameters of  propagation and survival dependencies between the species. We map the parametric set onto 
the basic types of space-time structures emerged in the mutualistic  population dynamic. The structures discovered
include propagating quasi-one dimensional patterns, very slowly growing clusters, still and oscillatory
stationary localizations. Although we hardly find such idealized patterns in nature, due to increased complexity of interaction phenomena, we recognize our findings as basic spatial patterns of mutualistic systems, which can be used as baseline to build up more complex models. 
\end{abstract}


\end{frontmatter}

\markboth{Adamatzky and Grube}{CA models of mutualism}

\section{Introduction}
\label{intro}

The term symbiosis was coined in 1879 by Heinrich Anton de Bary, a German mycologist, who defined it as the living together of unlike organisms\cite{debary_1879}. In biology, symbiosis includes all kinds of close relationships between species, ranging from parasitic interactions to mutualisms. In two-species interactions, mutualism is an interaction of species of organisms that benefits both~\cite{odum_1971}. Most classical symbioses were initially described as two-species interactions, but modern research shows that many of these cases comprise more interacting species and a higher level of complexity than previously thought~\cite{frey-klett_2007, little_2008}. In complex symbiotic systems, the distinction between beneficial and detrimental interactions among the partners is sometimes not clear or can be indirect. The outcomes of symbiotic associations may depend also on physiological circumstances, even when only two interacting partners are considered. Mutualism is the most intriguing and still not well understood type of inter-species interactions~\cite{bucher_1988, stadler_2008}. Even simplest models of mutualism exhibits higher behavioral complexity than predatoring, parasitism, amensalism and comensalism~\cite{adamatzky_populations}. In the present paper we decided to leave complex interactions and multispecies associations beyond the scope, focus on bipartite systems for the sake of simplicity. This simplification allows us to study idealized localization patterns to understand basic laws of spatial structuring without nuisance of real-life complexity.  

Mathematical and computer studies of mutualism are growing extensively last years. Most analytical models are based
on modifications of Lotka-Volterra model with positive inter-species interactions~\cite{gause_witt_1935,boucher_1985},
stabilized with feedbacks of limited resources~\cite{holland_2002}, tuned by diffusiveness and transport effects~\cite{delgado_1998}, and other relations between interacting species~\cite{neuhauser_2004,chen_2007}. Other mathematical
models are based on limit per capita growth~\cite{graves_2006} and feedback delays~\cite{liu_2008}.
 
Almost no results  are obtained in spatial evolution of mutualistic populations. Spatially extended 
prey-predator systems produce characteristic wave patterns, what are the patterns emerging in mutualistic
systems? So far only existence of stationary localized domains, or patches, of species is known. Their existence 
is demonstrated by two different 
techniques: lattice Lotka-Volterra model~\cite{tainaka} and reaction-diffusion model of population 
dynamics~\cite{morozov_2008}. Are patches the only patterns which could be observed in mutualistic populations? We aim to answer
the question in present paper. 

In computer simulations we employ cellular automata: two-dimensional arrays of final-state machines which update their states simultaneously and depending on states of their immediate neighbours.  Cellular automata have been used to simulate population dynamics for a long time. They become popular as population models from Dewdney articles on prey-predator systems~\cite{dewdney_1988}, and further supported by high-profile research on lattice-gas automata~\cite{camazine_1991} and automata models of host-parasite interaction ~\cite{hassell_1991}. For (dis)advantages of cellular automata models see 
overviews~\cite{ermentrout_1993} and~\cite{darwen_green_1996}. The automata models are now uncontested models for studying 
pattern formation in population dynamics~\cite{deutsch_2005}, pattern-oriented ecological modeling~\cite{grimm_1996} and spatial ecology~\cite{boccara_1994,tilman_1997,szaran_1997,dieckman_2000}.

Cellular automata 'substrates' are proved to be successful in imitating and simulating propagation of species~\cite{cannas_1999},
developments of plant populations~\cite{baltzer_1998}, prediction of epidemics dynamics in spatially heterogeneous 
environments~\cite{duryea_1999}, stochastic species invasion~\cite{cannas_2003,kizaki_1999}, predation chains~\cite{laan_1995} and competition in complex landscapes~\cite{caswell_1999}, and prey-predator systems~\cite{chen_2003}. Apart of particular case 
of interacting lattices~\cite{boza_2004} we are unaware of any published results concerting pure cellular automata models of mutualistic systems. A classical, in a sense of Ulam and von Neuman spirit, cellular automata model will be offered in present paper.

In the paper we utilise our ideas on automata-based modelling of population dynamics~\cite{adamatzky_1997}, designs of 
cell-state rules covering all types of inter-species interactions~\cite{adamatzky_1994}, and our scoping experiments on phenomenology of automata model of two-species populations~\cite{adamatzky_populations}.

The paper is structured as follows. In Sect.~\ref{model} we introduce a cell-state transitions rules, imitating mutualism.
Phenomenology of the models for various parameters of propagation and survival dependencies are outlined in Sect.~\ref{phenomenology}. Section~\ref{localizations} discusses basic types of localized patterns discovered and their 
parametric mappings.

\section{Automaton model of mutualism}
\label{model}

We study hexagonal lattice of finite-state machines, or cells, which take finite number of states and update their 
states in discrete time depending on states of their closest neighbours. Every cell $x$ 
has six neighbours, which determine the cell $x$'s neighbourhood $u(x)$ and takes 
three states: 0, 1 and 2. States 1 and 2 represent species `1' and species `2'. State 0 represent an `empty space', or
a substrate. 

Two processes must be simulated: propagation of species and survival of species. A cell of the hexagonal cellular 
automaton can be occupied exclusively by 'empty space', state 0, or by one of the species, states 1 or 2. 
A cell $x$ in `empty state' at time $t$ ($x^t=0$)  becomes occupied by species 1 at time $t+1$
if it has more neighbors in state 1 than in state 2 
($\sigma_1 > \sigma_2$, $\sigma_i = |\{ y \in u(x): y^t=i \}|$, $i \in \{1, 2 \}$) but 
enough neighbors in state 2 to support species 1 depending on them ($\sigma_2 > \theta_{01}$), or  it has equal amount of both species ($\sigma_1 = \sigma_2$) but there are more species 2 to support species 1 than species 1 to support species 2 
($\sigma_2 - \theta_{01} > \sigma_1 - \theta_{02}$). Propagation of species 2 can be discussed similarly. 

\begin{figure}
\centering
\begin{tabular}{c|ccc}
$x^t/x^{t+1}$ &  0                        &  1     & 2   \\ \hline
0            & $\sigma_1+\sigma_2=0$      &  $A \vee (\neg A \wedge B)$    & $C \vee (\neg C \wedge D)$      \\
1            & $\sigma_2<\theta_{11}$       &  $\sigma_2 \geq \theta_{11}$  &   never \\
2            & $\sigma_1<\theta_{22}$       &  never & $\sigma_1 \geq \theta_{22}$  \\
\end{tabular}
\caption{Cell-state transition table imitating mutualistic interactions. 
The conditions $A$, $B$, $C$ and $D$ are as follows:
$A = (\sigma_1 > \sigma_2) \wedge (\sigma_2 > \theta_{01}) \vee (\sigma_1 + \sigma_2>0)  \wedge (\sigma_2 - \theta_{01} > \sigma_1 - \theta_{02})$, 
$C = (\sigma_2 > \sigma_1) \wedge (\sigma_1 > \theta_{02}) \vee (\sigma_1 + \sigma_2 >0) \wedge (\sigma_2 - \theta_{01} < \sigma_1 - \theta_{02})$, 
$ B = ({\tt random}(1) < 0.5)$, 
$ D = ({\tt random}(1) \geq 0.5)$, and ${\tt random}$ generates a random real number between 0 and 1.
}
\label{transitiontable}
\end{figure}

A cell in state 1 (2) remains in state 1 (2) if number of its neighbors in state 2 (1) exceeds specified threshold, $\sigma_2 > \theta_{11}$ ($\sigma_1 > \theta_{22}$). There are no cell-state transitions $1 \rightarrow 2$ and $2 \rightarrow 1$. 
See summary of the cell-state transitions in Fig.~\ref{transitiontable}.

The thresholds $\theta_{01}$ and $\theta_{02}$ are propagation, or sustainability parameters. The thresholds 
$\theta_{11}$ and $\theta_{22}$ are survivability parameters. 
In general, $1 \leq \theta_{01}, \theta_{02}, \theta_{11}, \theta_{22} \leq 6$, however no activity persists on the lattice 
for values higher than 3, so in the paper we consider only $1 \leq \theta_{01}, \theta_{02}, \theta_{11}, \theta_{22} \leq 3$.

Sometimes we are addressing a cell-state rule as rule $R(\theta_{01}, \theta_{02}, \theta_{11}, \theta_{22})$.

\section{Phenomenology}
\label{phenomenology}

\graphicspath{{mutual_v_05/}}
\begin{figure}
\centering
 \includegraphics[width=0.9\textwidth]{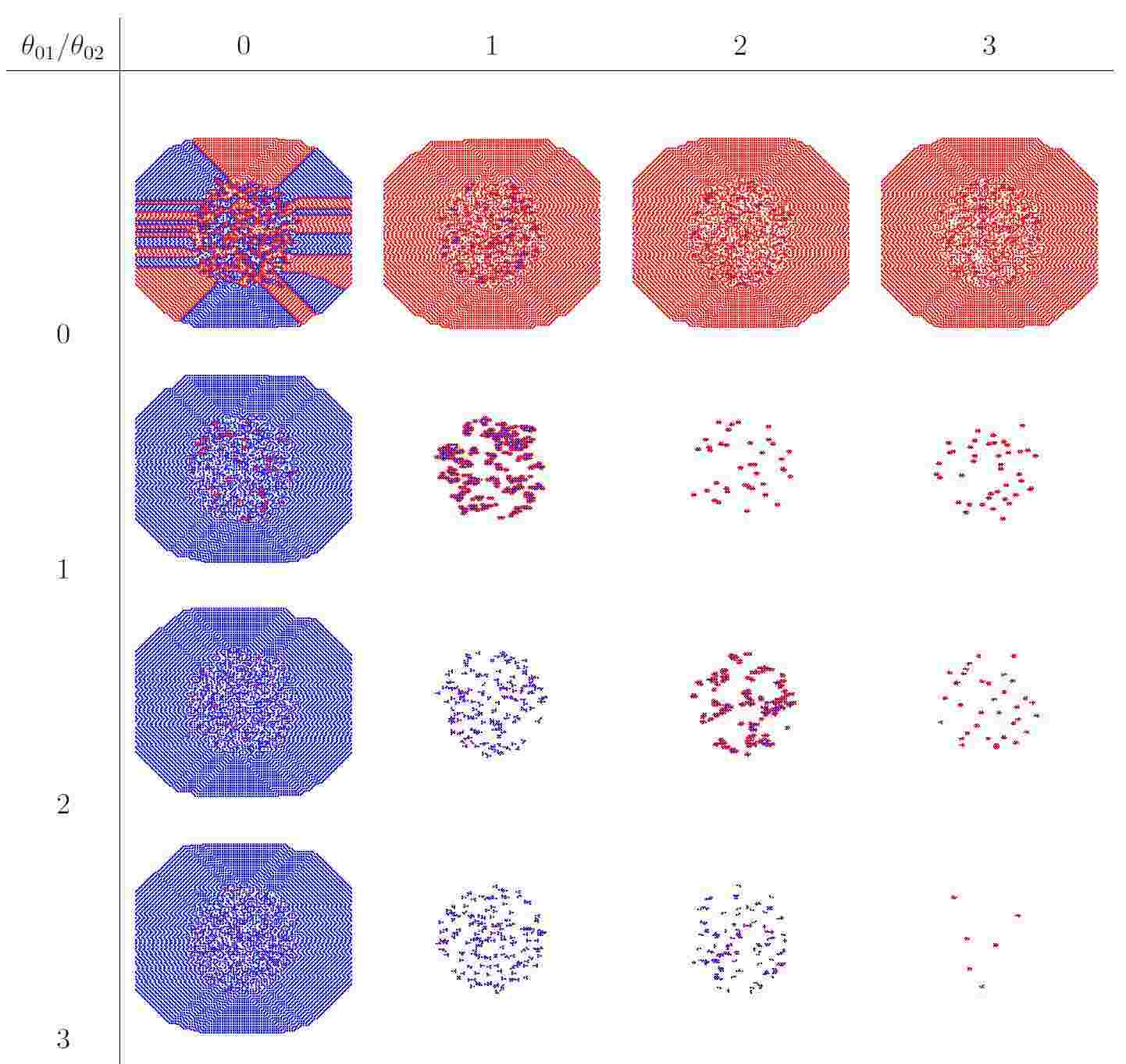}
\caption{Samples of configurations parameterized by values of $\theta_{01}$, rows, and $\theta_{02}$, columns. 
Each entry $[\alpha, \beta]$ in the table represents a sample configuration for parameters 
$\theta_{01}=\alpha$, $\theta_{02}=\beta$, $\theta_{11}=1$ and $\theta_{22}=2$. Lattice is $200 \times 200$. 
Initially cells not further then 50 cells from the lattice centre are assigned states 1 or 2 with probability
0.1 each.}
\label{figSamples}
\end{figure}

We have analysed development of automata from random initial configurations for all values of four 
thresholds $1 \leq \theta_{01}, \theta_{02}, \theta_{11}, \theta_{22} \leq 3$. We found that 
when one of the species 1 (2) does not depend on another species to propagate $\theta_{02}=0$ 
($\theta_{01}=0$) then this species 1 (2) propagates unlimitedly while other species remains  
confined to the zone of initial inoculation. See comprehensive list of configurations in Appendix
and few selected samples in Fig.~\ref{figSamples}.
 
Dynamics of unlimited propagation depends on parameters $\theta{01}$ and $\theta_{02}$. 
Consider for example, $\theta_{01}=0$ and $\theta_{02}<4$. Species $1$ propagates on the 
lattice because they do 
not need species $2$ to invade new space. If $\theta_{11}=0$ then lattice becomes filled with `solid' 
pattern of species $1$, once a site occupied by $1$ it always remains in the state $1$. 
For $\theta_{11}>0$ we observe structures similar to target waves because species $1$ can not 
remain at the same site for more then one step of discrete time $t$, it dies, or vacates the site, 
at time $t+1$, however the site becomes occupied again by species $1$ at time step $t+1$. 

For the parameters $\theta_{02}>0$ and $\theta_{01}>0$ localizations --- compact 
groups of non-quiescent (not equal 0) states --- are observed. The number of localizations 
in configurations, recorded after transient period, decreases with increase 
of $\theta_{02}+\theta_{01}$. Principle localizations are discussed in next section.

\section{Localizations in mutualistic systems}
\label{localizations}

In exhaustive experiments we discovered four principle types of localizations: stationary still localizations,
stationary oscillating localizations, propagating quasi-one-dimensional localizations, or worms, and 
sub-linearly growing domains. `Stationary' means that the pattern does not travel along the lattice, 
`still' means that the pattern does not change its structure during automaton development.

\begin{figure}
\centering
\includegraphics[width=0.4\textwidth]{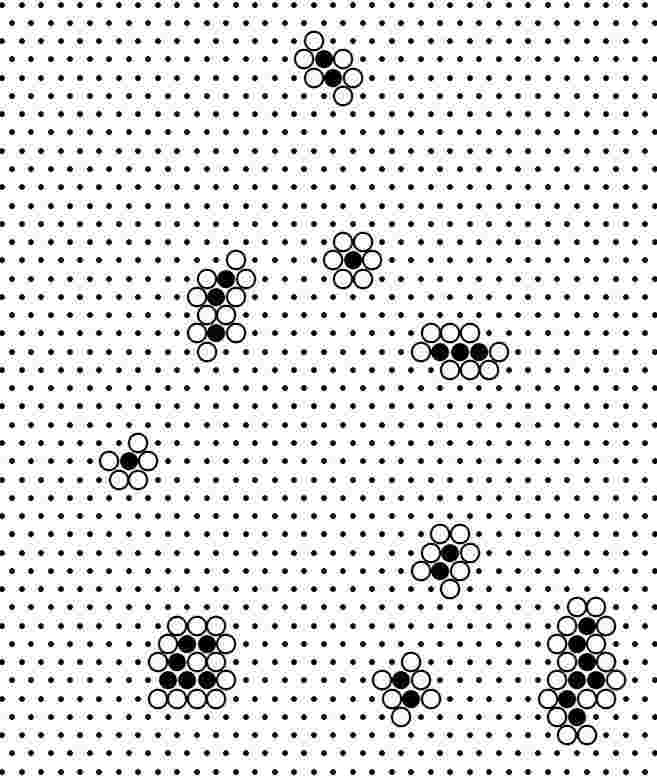}
\caption{Stationary still localizations in rule $R(2221)$. State `1' is shown by solid disc,
state `2' by circle, state `0' by dot.}
\label{still2221}
\end{figure}

Stationary still localization is a compact groups of cells in non-quiescent (non-`0') states. Diameter of such group
usually does not exceed 1-3 cells. Typically one species occupies central position in the localization and
it is surrounded by sites occupied by another species. Thus in Fig.~\ref{still2221} we see sites of 
species 1 (solid discs) surrounded by several sites with species 2 (circles). Species 1 survive because they 
have more than two neighbouring sites with species 2 ($\theta_{02} > 2$) while less dependent species 2 are 
supported by being in relation with just one site occupied by species 1 ($\theta_{01} > 1$).  Several elementary
localizations can be linked together to form an extended cluster of localizations.

\begin{figure}
\centering
\includegraphics[width=0.9\textwidth]{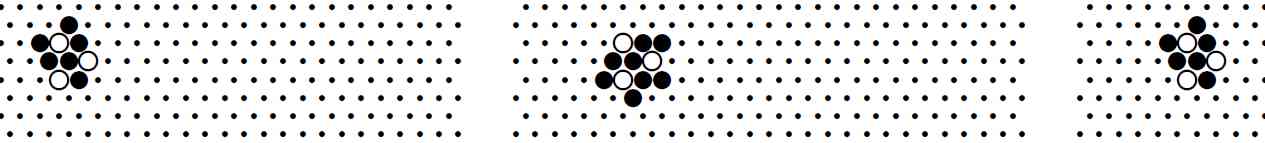}
\caption{Oscillators in rule $R(2323)$.}
\label{oscil2323}
\end{figure}

\begin{figure}
\centering
\subfigure[$t$]{\includegraphics[width=0.4\textwidth]{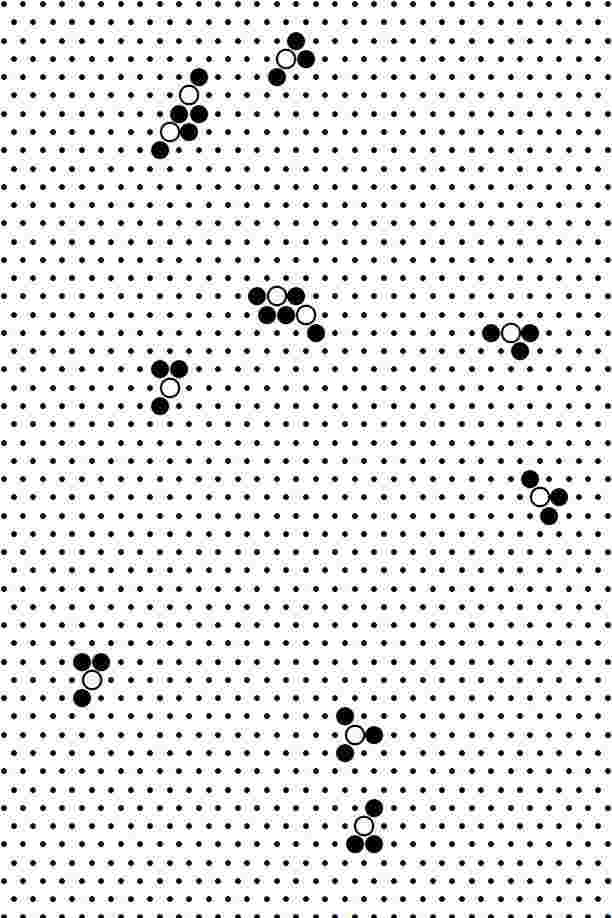}}
\subfigure[$t+1$]{\includegraphics[width=0.4\textwidth]{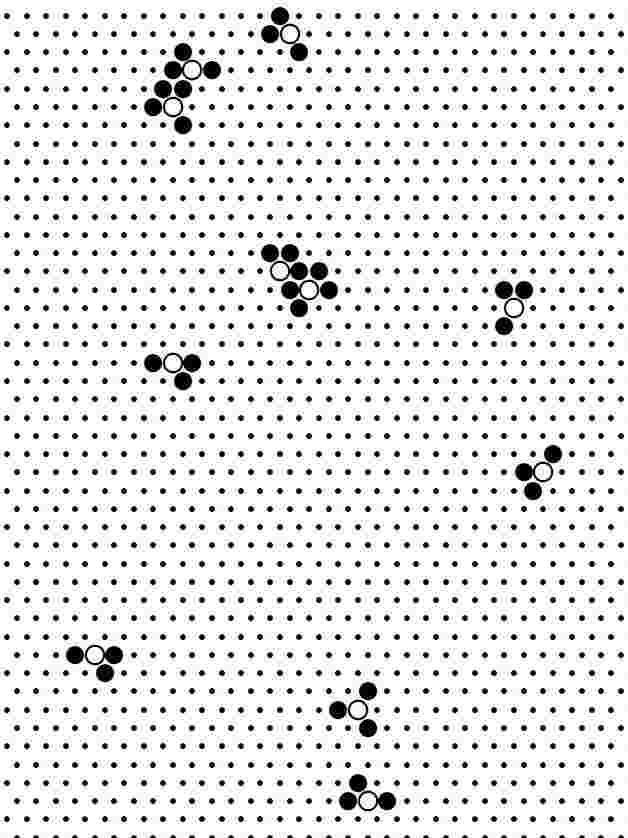}}
\caption{Oscillators in rule $R(1223)$.}
\label{oscill1223}
\end{figure}

\begin{figure}
\centering
\subfigure[$t$]{\input{locals1113/0058}}
\subfigure[$t+1$]{\input{locals1113/0059}}
\subfigure[$t+2$]{\input{locals1113/0060}}
\subfigure[]{\includegraphics[width=0.69\textwidth]{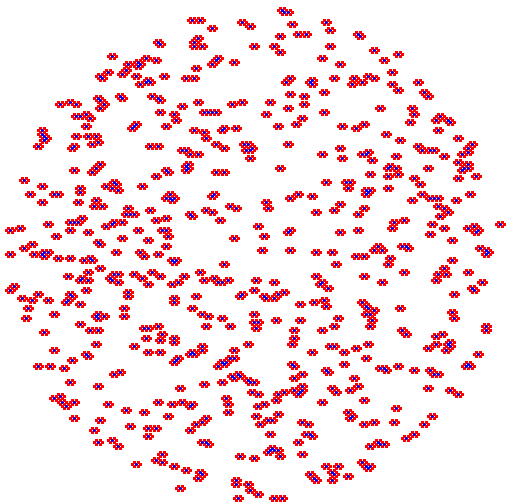}}
\caption{Still localizations and oscillators in  rule $R(1113)$. (a)-(c)~three consecutive 
configurations exhibiting still localizations: sites with species 2 surrounded by sites with species 1, 
and stationary oscillators: sites with species surrounded by sites with species 2. 
(d)~zoomed out configuration.
}
\label{1113}
\end{figure}

Stationary oscillator is compact group of cells in non-quiescent states, which does not move along the lattice
unlimitedly. The group can be translated cyclically around some fixed point. 
All oscillators observed in space-time dynamics of mutualistic populations are can be classified as flip-flops, which
rotate on a fixed degree around stationary centre, e.g. oscillators in Fig.~\ref{oscil2323} and Fig.~\ref{oscill1223}, 
and breathing oscillators, which has stationary core, e.g. sites in state 1 in Fig.~\ref{1113}, and 
switching/breathing halo, e.g. sites in state 2 in Fig.~\ref{1113}.

Some rules support only still localizations, e.g. rule $R(2221)$ (Fig.~\ref{still2221}), 
some only oscillators, e.g. rule $R(2323)$ (Fig.~\ref{oscil2323}), 
others support both still and oscillating localizations, 
e.g. rule $R(1113)$ (Fig.~\ref{1113}).

\begin{figure}
\centering
\includegraphics[width=0.9\textwidth]{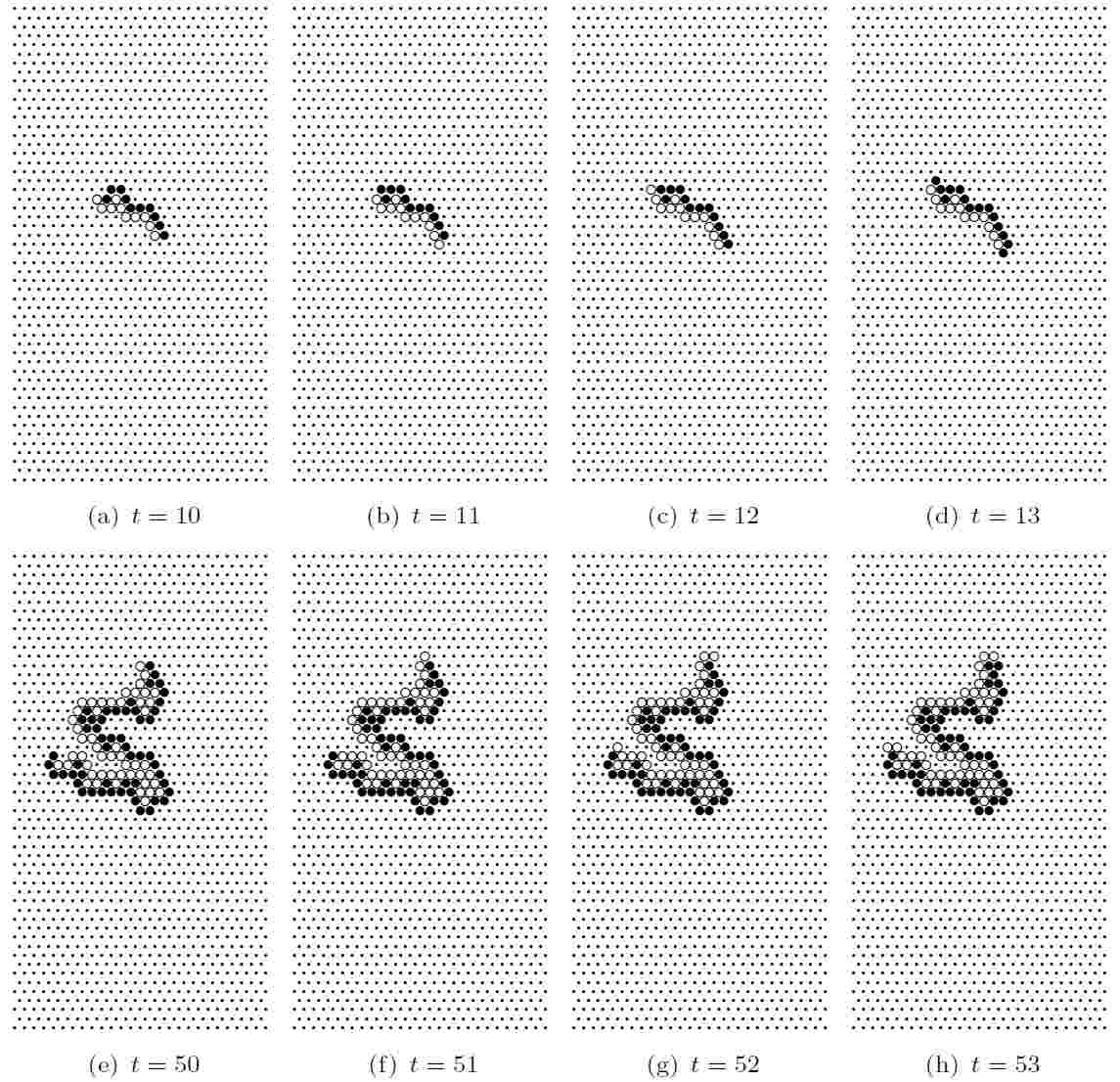}
\caption{Propagation of long-life worms in rule $R(1111)$.}
\label{worms1111A}
\end{figure}

\begin{figure}
\centering
\subfigure[]{\input{texfigsworms1/0197}}
\subfigure[]{\includegraphics[width=0.49\textwidth]{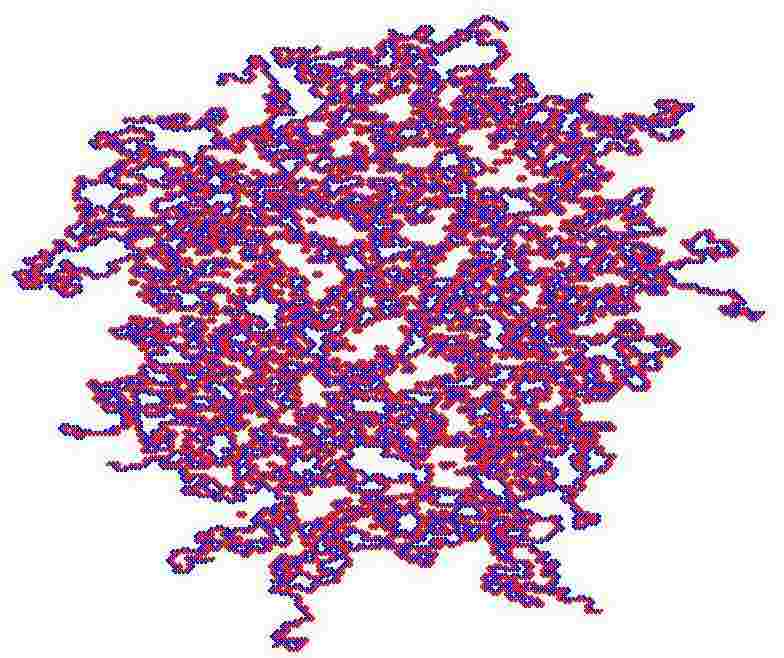}}
\caption{Long-life worms in rule $R(1111)$. (a)~a single worm grown for 197 steps of discrete time.
(b)~Zoomed out configuration of rule $R(1111)$ automaton, which starts its initial development in a random configuration, where
all cells closer then 150 sites to the centre of the lattice are assigned states 1 or 2 with probability 0.5.}
\label{worms1111B}
\end{figure}

Worms are quasi-one-dimensional propagating localizations, which usually consists of two parallel 
chains of non-quiescent states, one chain is formed by sites with species 1, another chain by 
sites with species 2. All worms observed in our experiments have only two growing tips. For example, 
see in Fig.~\ref{worms1111A} snapshots of a single worm growing in rule $R(1111)$ automaton. The worm 
has two growing tips which are two sites: one occupied by state 1, another by state 2. For neighbouring
quiescent cells numbers of non-quiescent states 1 and 2 is equal, so the states 1 and 2 `diffuse' randomly. 
Species 1 and 2 can not survive without each other, therefore when one of the species advances, it must then
`wait' for a neighbouring empty site to be occupied by another species. Due to randomization involved the 
growing tips do not propagate directly but rather implement a kind of random walk on the lattice (Fig~\ref{worms1111B}a).

When initial density of species 1 and 2 is high enough, many worms are born (Fig~\ref{worms1111B}b), 
they propagate and collide with each other. When a growing tip collides to a body of a worm it stops 
propagating. Thus forming a porous stationary core with free growing tips propagating 
centrifugally~(Fig~\ref{worms1111B}b).  

\begin{figure}
\centering
\includegraphics[width=0.9\textwidth]{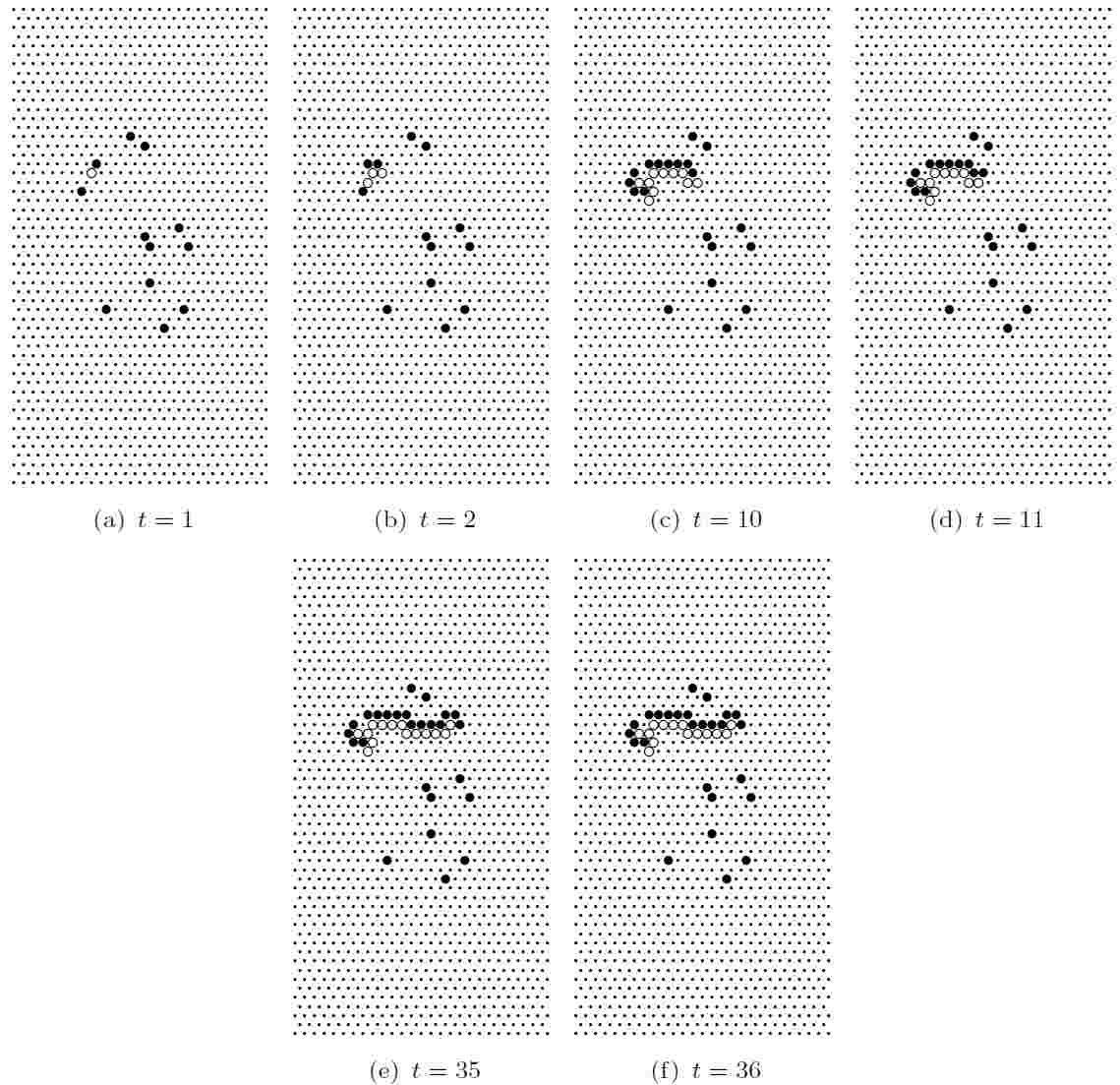}
\caption{Propagation of short-life worms in rule $R(3300)$.}
\label{worms3300}
\end{figure}

For some rules, e.g. rule $R(3300)$ (Fig.~\ref{worms3300}), worms are short-living.  In these rules worms 
stop propagating when their growing tips are blocked by the following conditions. Every resting cell $x$, $x^t=0$, 
closest to non-resting cells of the worm's tip has $(\sigma^t_1 > \sigma^t_2) \wedge  (\sigma^t_2 < \theta_{01})$, i.e. 
species 1 do not propagate, and also $(\sigma^t_2 > \sigma^t_1) \wedge  (\sigma^t_1 < \theta_{02})$, i.e. species 2 do not propagate as well.

\clearpage 
\
\begin{figure}
\centering
\includegraphics[width=0.79\textwidth]{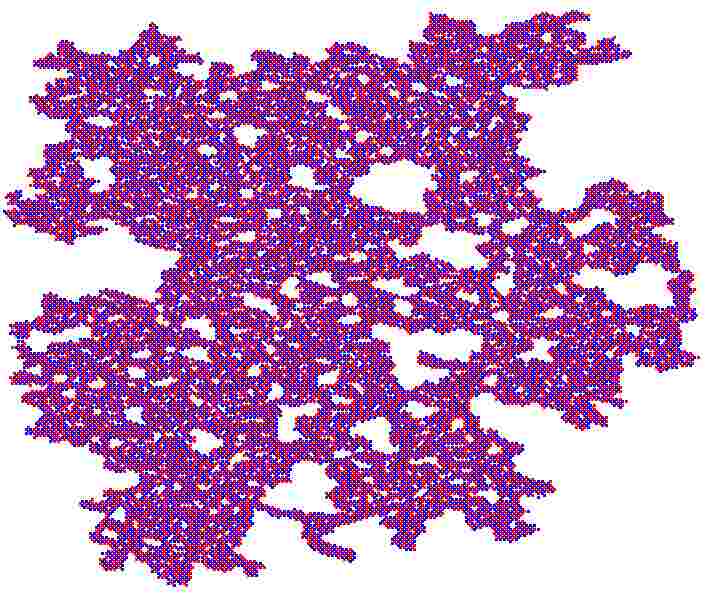}
\caption{A snapshot of a sub-linearly growing stripy domains, rule $R(2222)$, taken at 1171th step of automaton development from 
initial random configuration.}
\label{strippy2222}
\end{figure}

The last class of localizations are very slowly growing stripy domains (Fig.~\ref{strippy2222}), found in space-time 
dynamics of cellular automata governed by rules $R(2222)$ and $R(3322)$. They are not proper localizations, because their size is constantly increasing. However it does increase very slow: from computational experiments we estimated that a radius of a circle around a stripy domain grows as $0.05 \cdot t$, where $t$ is a number of time steps from the beginning of the pattern development. That is the growth rate is almost two orders smaller then typical propagation rate of usual cellular automaton patterns (e.g. excitation waves or gliders in Conway's Game of Life). So comparing to typical cellular automaton growing patterns the stripy domains behave as localizations.

\begin{figure}
\centering
\includegraphics[width=0.9\textwidth]{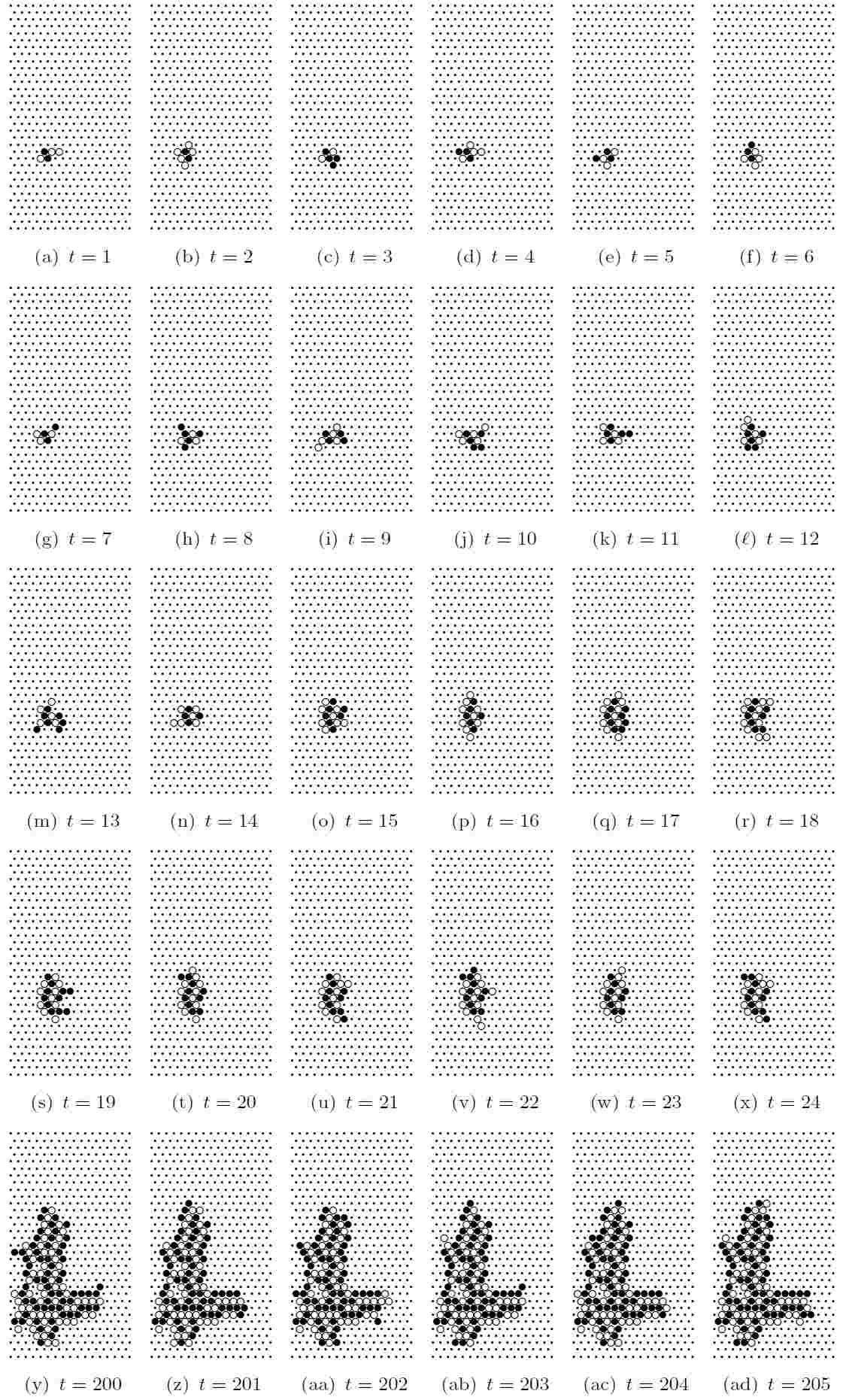}
\caption{Mechanics of stripy pattern formation in  rule $R(2222)$.}
\label{stripyTex}
\end{figure}

The domains are stripy, i.e. consist of intermittent chains of 1 and 2 states, because each species needs at least two or three sites occupied by other species to propagate and at least two sites occupied by other species to survive. Thus having a chain of
states 1, the chain will be covered by on both sides by states 2, which will be covered by states 1 and so on. See mechanics of 
the pattern formation in Fig.~\ref{stripyTex}. Growth of the patterns is slow due to involved of randomization, see conditions
$B$ and $C$ in cell-state transition table (Fig.~\ref{transitiontable}). When an empty site accept species 1 or 2 at random, the species may not stay there if survivability criteria not met, then the site become empty again and may take of the species at random. For example, see how species 2 propagate to the North tip of the growing pattern in Fig.~\ref{stripyTex}v--x. At first 
empty site is occupied by species 2 (also because there are neighbouring species 1), however species 1 disappear because their
neighbourhoods do not meet survivability criteria; and the site just occupied by species 1 becomes empty again (Fig.~\ref{stripyTex}x). 
 
\section{Conclusions}

In the paper we introduced a minimal cellular automaton model of a two-species mutualism, where every cell of a two-dimensional 
hexagonal lattice takes three states: species 1 and 2 and empty space 0, and updates its state depending on number of another species in its neighbourhood. The cell-state transition rules are threshold-based: a species propagate if number of another species exceeds certain threshold (propagation dependence) and the species continue occupying the newly acquired site if number of another
species exceed certain threshold (survivability dependence). We were mainly concerned with a particular type of mutualism when 
both species can not propagate without each other but can survive without each other.

We have discovered four types of localizations --- loci of non-empty sites which either preserve their size or grow very slowly. They are stationary still localizations, stationary oscillators, quasi-one-dimensional propagating domains and 
sub-linearly growing stripy domains.

We were looking in the natural world for localization which resemble the ones presented in this work. However, we were unable to find any of the patterns in natural symbiotic systems. Evidence for spatial self-organization is present in mussel communities. Small-scale, labyrinthine spatial patterning, which resemble somehow our sub-linearly growing stripy domains, are present in beds of the
blue mussel \textit{Mytilus edulis}~\cite{koppel_2008}. However, these patterns are not the results of symbiotic interactions. Previous work showed that an empirically derived cellular automaton approach could predict macroscale patterns~\cite{wootton_2001}. The rather complex full cellular automaton model comprised 15 ecological states which also include states for co-occurring other species. The spatial structures formed by this model are patchy clusters of mussels with separating gaps, with their size distributions being dependent on the inclusion of interactions with other species. 

Why do we not find the idealized patterns in nature ? Even the simplest cases of inter-species interactions are apparently more complex. Many species produce diffusible compounds which can have signalling or nutritional function, blurring the local spatial consequences of interactions among the species, which and can change the behaviour of species, such as exemplified by quorum-sensing. Colonization of natural substrates is usually not restricted to two dimensions. In early or late stages bacterial multispecies biofilms grow in the third dimension to produce characteristic shapes and phenomena, which are often associated with altruistic production of extracellular polysaccharides in one or more of the interacting species~\cite{xavier_2006}. In addition, the substrate is often not exactly homogeneous, and even minute variations of surfaces can cause responses of the emerging community. Finally, natural symbionts can be motile to translocate the partner species. However, our simplified model represents principal rules of spatial patterning in mutualisms which may serve as starting points to buid up more complex models of associations.

\begin{figure}
\centering
\begin{tiny}
\begin{tabular}{p{0.05\textwidth}|p{0.3\textwidth}p{0.3\textwidth}p{0.3\textwidth}}
 $\theta_{01}$/$\theta_{02}$  & 1  & 2 & 3 \\  \hline
                  1           & Propagating worms: $R(1100)$, $R(1101)$, $R(1111)$; 
                                Very slowly-growing worms/domains: $R(1122)$;
                                Still and oscillating domains for other values of 
                                parameters.  &  
                                All clusters are non-propagating. 
                                Majority are stationary still clusters. 
                                Only for $R(1222)$, $R(1223)$, $R(1233)$ oscillating clusters. &  
                                Mostly stationary still clusters, 
                                oscillating clusters only for $R(1322)$, $R(1323)$, $R(1333)$.  \\  
                  2           & --- & Worms propagate but freeze quickly: $R(2200)$, $R(2201)$, $R(2202)$;
                                      For $R(2202)$ life-time of worms  is really short and 
                                      no worms appear for $R(2203)$; for $R(2211)$ worms freeze at once 
                                      and form still clusters. 
                                      For $R(2212)$ and $R(2213)$ still clusters,  $R(2222)$ 
                                      slowly growing stripy domains. 
                              &  Mostly still stationary clusters, oscillating only
                                 for $R(2322)$, $R(2323)$, for $R(2333)$ all activity extinguish all dying.\\
                  3           &  ---    &  ---  &   For $R(3300)$ worms propagate but freeze quickly; 
                                                    for $R(3301)$--$R(3303)$ number and life-time of warms 
                                                    decrease, till no worms at all for $R(3303)$;
                                                    $R(3311)$ long-living worms; for $R(3312)$ to $R(3313)$ 
                                                    worms quickly freeze or die. For $R(3322)$ worm-like
                                                    large clusters grow. There is no sustained activity for 
                                                    $R(3323)$ and $R(3333)$ no sustained
                                                    activity. \\                                    
\end{tabular}
\end{tiny}
\caption{Characteristics of space-time dynamics for $\theta_{01}>0$, rows, and $\theta_{02}>0$, columns. 
Each entry $[\alpha, \beta]$ provides a brief description of space-time dynamics of  
a sample configuration for parameters $\theta_{01}=\alpha$, $\theta_{02}=\beta$, $0 \leq \theta_{11} \leq 3$ and 
$0 \leq \theta_{22}=2 \leq 3$.}
\label{characterisation}
\end{figure}

Space-time dynamic of mutualistic populations parameterized by propagation dependencies is shown 
in Fig.~\ref{characterisation}. There we see that when propagation dependencies increase or when 
dependence of one species is higher then dependence of another species the stationary localizations
change from propagating worms to oscillation localizations to stationary localizations. Most remarkable
varieties of localizations emerge for lower but above zero propagation dependencies. The very similar 
correlation takes place between spatial patterns and survivability parameters (see full catalog of patterns in Appendix), i.e. 
transitions from worms to oscillators to still localizations.



\begin{thebibliography}{99}

\bibitem{adamatzky_1997}
Adamatzky A. Cellular automaton labyrinths and solution finding,
Computers \& Graphics 21 (1997) 519--522.

\bibitem{adamatzky_1994}
Adamatzky A. Identification of Cellular Automata, Taylor \& Francis, 1994. 

\bibitem{adamatzky_populations}
 Adamatzky~A. Minimal model of two-species interactions on
 hexagonal lattice (2008), submitted.
 
 \bibitem{baltzer_1998}
Balzter H., Braun P.W. and K\"{o}hler P.
Cellular automata models for vegetation dynamics
Ecological Modelling 107 (1998) 113--125.

\bibitem{debary_1879}
de Bary A.
Die Erscheinungen der Symbiose
Strassburg: Trübner, 1879.  30 S.

 \bibitem{boccara_1994}
Boccara N. 
Automata network models of interacting population. In: 
Goles E. and Martinez S. (Eds.) 
Cellular Automata, Dynamical Systems and Neural Networks. 
Springer, 1994.
 
 \bibitem{boucher_1985}
Boucher~D.~H. 
The idea of mutualism, past and future. In:
Boucher~D.~H. (Ed.), The Biology of Mutualism 
(Oxford University Press, New York, 1985), 1–-28. 

\bibitem{bucher_1988}
Boucher~D.~H.
The Biology of Mutualism: Ecology and Evolution.
Oxford University Press, 1988.
 


\bibitem{boza_2004}
Boza~G., Scheuring~I.
Environmental heterogeneity and the evolution of mutualism.
Ecological Complexity 1 (2004) 329–-339.

\bibitem{camazine_1991}
Camazine S.
Self-organizing pattern formation on the combs of honey bee colonies. Behav. Ecol. Sociobiol. 28 (1991) 61-–76.

\bibitem{cannas_1999}
Cannas S.A., P\'{a}ez S.A. and Marco D.E.
Modeling plant spread in forest ecology using cellular automata
Computer Physics Communications, 121/122 (1999) 131--135.

\bibitem{cannas_2003}
Cannas S.A., Marco D.E. and P\'{a}ez S.A. 
Modelling biological invasions: species traits, species interactions, and habitat heterogeneity
Mathematical Biosciences 183 (2003) 93--110.
 
 
\bibitem{caswell_1999}
Caswell H. and Etter R.
Cellular automaton models for competition in patchy environments: Facilitation, inhibition, and tolerance,
Bulletin of Mathematical Biology 61 (1999) 625-649.


\bibitem{chen_2003}
Chen Q. and Mynett A.E.
Effects of cell size and configuration in cellular automata based prey–predator modelling
Simulation Modelling Practice and Theory 11 (2003) 609--625.

\bibitem{chen_2007}
Chen~F. and You~M.
Permanence for an integrodifferential model of mutualism.
Applied Mathematics and Computation 186 (2007) 30–-34.

 
 \bibitem{darwen_green_1996}
Darwen P.J. and D. G. Green, Viability of populations in a landscape
Ecological Modelling 85 (1996) 165--171.

\bibitem{delgado_1998}
Delgado~M.,  L\'{o}pez-Go\'{o}mez~J. and Su\'{a}rez~A.
On the symbiotic Lotka-Volterra model with diffusion
and transport effects.
Journal of Differential Equations 160 (2000) 175--262.

\bibitem{deutsch_2005}
Deutsch A. and Dormann S. 
Cellular Automaton Modeling of Biological Pattern Formation.
Birkh\"{a}user, 2006.

 
 \bibitem{dewdney_1988}
Dewdney A. K. 
Armchair Universe: An Exploration of Computer Worlds 
(Freeman and Co, 1988).

\bibitem{dieckman_2000}
Dieckmann U., Law R., Metz J.A.J. 
The Geometry of Ecological Interactions: Simplifying Spatial Complexity. 
Cambridge University Press, 2000.


\bibitem{duryea_1999}
Duryea M., Caraco T., Gardner G., Maniatty W. and Szymanski B.K.
Population dispersion and equilibrium infection frequency in a spatial epidemic
Physica D 132 (1999) 511-519.

\bibitem{ermentrout_1993}
Ermentrout G.B. and L. Edelstein-Keshet, Cellular automata approaches to biological modeling
J. Theor. Biology 160 (1993) 97--133.
 
\bibitem{frey-klett_2007}
Frey-Klett P., Garbaye J., Tarkka, M.
The mycorrhiza helper bacteria revisited.
New Phytologist 176 (2007) 22--36.

 \bibitem{gause_witt_1935}
Gause~G.~F. and Witt~A.~A.
Behavior of mixed populations and the problem of natural selection.
The American Naturalist 69 (1935) 596--609.


\bibitem{graves_2006}
Graves~V.G., Peckham~B.~B., Pastor~J.
A 2D differential equations model for mutualism.
Department of Mathematics and Statistics.
Technical Report TR 2006-2. University of Minnesota Duluth
2006.



\bibitem{grimm_1996}
Grimm V., Frank K., Jeltsch F., Brandl R., Uchmanski J. and Wissel C.
Pattern-oriented modelling in population ecology, 
Sci. of The Total Environment 183 (1996) 151--166.


\bibitem{hassell_1991}
Hassell M. P., S. W. Pacala, R.M. May and  P.L. Chesson 
The persistence of host-parasitoid associations in patchy environments. I. A general criterion. American Naturalist 138 (1991) 568--583.


\bibitem{holland_2002}
Holland~J.~N., DeAngelis~D.~L., Bronstein~J.~L.
Population dynamics and mutualism: functional responses of benefits and
costs. American Naturalist 159 (2002) 231–-244.



\bibitem{kizaki_1999}
Kizaki S. and Katori M.
A stochastic lattice model for locust outbreak
Physica A 266 (1999) 339--342.

\bibitem{koppel_2008}
van de Koppel J., Gascoigne J.C., Theraulaz G., Rietkerk M., Mooij W.M., Herman P.M.J. 
Experimental Evidence for Spatial Self-Organization and Its Emergent Effects in Mussel Bed Ecosystems. 
Science 322 (2008) 937--942.

\bibitem{laan_1995}
van der Laan J.D., Lhotka L. and Hogeweg P.
Sequential predation: a multi-model study
Journal of Theoretical Biology 174 (1995) 149-167.

\bibitem{little_2008}
Little A.E.F. and Currie C.R.
Black yeast symbionts compromise the efficiency of antibiotic defenses in fungus-growing ants.
Ecology 89 (2008) 1216--1222.

\bibitem{liu_2008}
Liu~Z., Tan~R., Chen~Y., Chen~L.
On the stable periodic solutions of a delayed two-species
model of facultative mutualism.
Applied Mathematics and Computation 196 (2008) 105–-117.

\bibitem{morozov_2008}
Morozov~A., Ruan~S. and Li~B.-L.
Patterns of patchy spread in multi-species
reaction–diffusion models.
Ecological Complexity (2008).


\bibitem{neuhauser_2004}
Neuhauser~C. and Fargione~J.~E.
A mutualism-parasitism continuum model and its application
to plant-mycorrhizae interactions. 
Ecological Modelling 177 (2004) 337–-352.


\bibitem{odum_1971}
Odum~E.~P.
Fundamentals of Ecology(W B Saunders Co, New York, 1971).


\bibitem{stadler_2008}
Stadler~B. and Dixon~A.~F.~G. 
Mutualism: Ants and their Insect Partners.
Cambridge University Press, 2008.


\bibitem{szaran_1997}
Szaran T. 
Spatiotemporal Models of Population and Community Dynamics. 
Springer, 1997.

\bibitem{tainaka}
Tainaka~K., Terazawa~N., Yoshida~N., Nakagiri~N., Takeuchi~Y.
Spatial pattern formation in a model ecosystem:
exchange between symbiosis and competition.
Physics Letters A 282 (2001) 373–-379.

\bibitem{tilman_1997}
Tilman D. and Kareiva P. (Eds.)
Spatial Ecology. Princeton University Press, 1997.

\bibitem{xavier_2006}
Xavier J.B. and Foster K.R.
Cooperation and conflict in microbial biofilms.
PNAS 104 (2006) 876--881.

\bibitem{wootton_2001}
Wootton J.T. 
Local interactions predict large-scale pattern in empirically derived cellular automata. 
Nature 413 (2001) 841--843.

\bibitem{wright_1989}
Wright~D.~H.
A simple, state model of mutualism incorporating
handling time. American Naturalist 134 (1989) 664–-667.


\end{thebibliography}
\end{document}